\documentclass{emulateapj}
\usepackage{apjfonts}

\newcommand{\gapprox}{\mathrel{\mathpalette\@versim>}}
\newcommand{\lapprox}{\mathrel{\mathpalette\@versim<}}
\newcommand{\propapprox}{\mathrel{\mathpalette\@versim\propto}}

\shorttitle{Nonthermal emission in G330.2+1.0} 
\shortauthors{WILLIAMS ET AL.}

\begin{document}

\title{A Deep X-ray View of the Synchrotron-Dominated Supernova Remnant G330.2+1.0}

\author{Brian J. Williams,\altaffilmark{1}
John W. Hewitt,\altaffilmark{2}
Robert Petre,\altaffilmark{3}
Tea Temim,\altaffilmark{1}
}

\altaffiltext{1}{Space Telescope Science Institute, Baltimore, MD 21218, USA}
\altaffiltext{2}{University of North Florida, Jacksonville, FL, USA}
\altaffiltext{3}{NASA Goddard Space Flight Center, Greenbelt, MD 20771}

\begin{abstract}

We present moderately deep (125 ks) {\it XMM-Newton} observations of supernova remnant G330.2$+$1.0. This remnant is one of only a few known that fall into "synchrotron-dominated" category, with the emission almost entirely dominated by a nonthermal continuum. Previous X-ray observations could only characterize the spectra of a few regions. Here, we examine the spectra from fourteen regions surrounding the entire rim, finding that the spectral properties of the nonthermal emission do not vary significantly in any systematic way from one part of the forward shock to another, unlike several other remnants of this class. We confirm earlier findings that the power-law index, $\Gamma$, ranges from about 2.1-2.5, while the absorbing column density is generally between 2.0-2.6 $\times 10^{22}$ cm$^{-2}$. Fits with the {\it srcut} model find values of the roll-off frequency in the range of 10$^{17.1} - 10^{17.5}$ Hz, implying energies of accelerated electrons of $\sim 100$ TeV. These values imply a high shock velocity of $\sim 4600$ km s$^{-1}$, favoring a young age of the remnant. Diffuse emission from the interior is nonthermal in origin as well, and fits to these regions yield similar values to those along the rim, also implying a young age. Thermal emission is present in the east, and the spectrum is consistent with a $\sim 650$ km s$^{-1}$ shock wave encountering interstellar or circumstellar material with a density of $\sim 1$ cm$^{-3}$.


\end{abstract}

\section{Introduction}
\label{intro}

It is well-known that shockwaves in supernova remnants (SNRs) accelerate some particles to nonthermal energies, well beyond the few keV thermal energies that the majority of the ions and electrons are heated to. While the particles in the thermal regime are responsible for the bulk of the X-ray emission, both in the form of lines and bremsstrahlung continuum, the nonthermal electrons, accelerated to relativistic energies of tens of TeV, spiral in the post-shock magnetic field, creating synchrotron emission. In general, shock velocities must be over a few thousand km s$^{-1}$ to produce synchrotron emission, though there are some exceptions \citep{williams11}.

While nonthermal synchrotron emission has been observed in perhaps a few dozen remnants in the Galaxy, there are only five remnants that are entirely dominated by this emission: G330.2+1.0 (the subject of this paper), G353.6$-$0.7, RX J0852.0-4622 (a.k.a., Vela Jr.), RX J1713.7-3946, and G1.9+0.3. The designation of "synchrotron-dominated" is not 100\% accurate, as some of these remnants, such as G1.9+0.3 and RX J1713.7-3946 do show hints of faint thermal emission as well \citep{borkowski13,katsuda15}. Additionally, the remnant of SN 1006 is often included in this class in the literature, despite having significant thermal emission from both the ejecta in the center of the remnant and the shocked interstellar medium (ISM) in the NW quadrant. However one counts the members of this class, there is no question that this small group of SNRs are important laboratories for studying the physics of electron acceleration to nonthermal energies in shock waves. In this paper, we report on moderately deep {\it XMM-Newton} observations of G330.2+1.0. We discuss the spatial variations of the X-ray spectra, and report on the confirmed presence of thermal emission in the eastern lobe of the remnant, as first reported in (\citealt{park09}, hereafter P09).

\begin{figure}[htb]
\includegraphics[width=8cm]{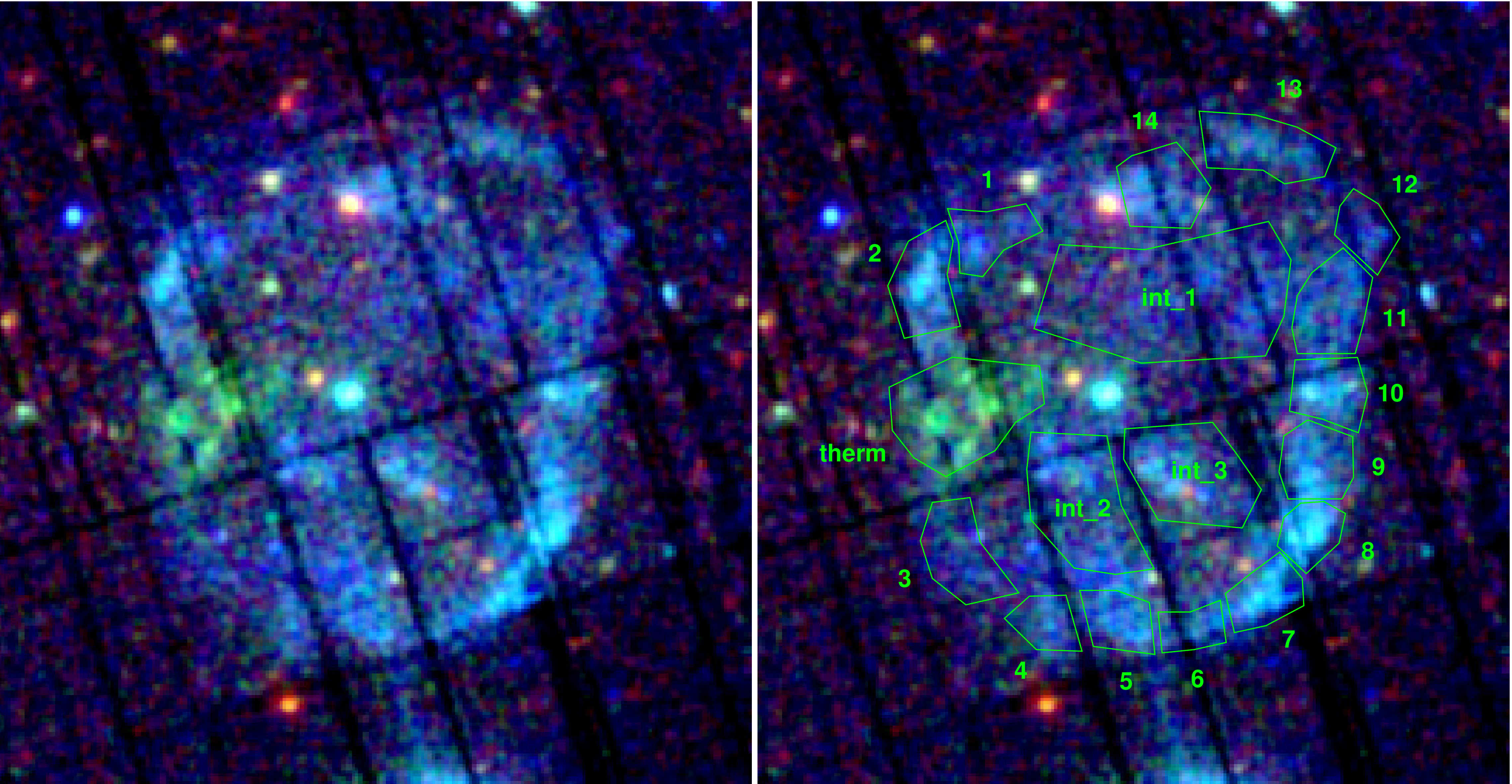}
\caption{Left: {\it XMM-Newton} three-color image of G330.2+1.0, with 0.4$-$1.2 keV emission in red, 1.2$-$2.0 keV emission in green, and 2.0$-$7.0 keV emission in blue. The image is 15$'$ wide. Right: Same image, but with our analysis regions overlaid.
\label{images}
}
\end{figure}

G330.2+1.0 is the remnant of a core-collapse supernova, as evidenced by the bright central compact object (CCO), CXOU J160103.1$-$513353. The remnant is moderately absorbed, with an N$_{H}$ value of $\sim 2.5 \times$ 10$^{22}$ cm$^{-2}$. The morphology is shell-like in the X-rays, though the shape of the shell is somewhat elliptical, roughly 12.5$'$ $\times$ 10.5$'$ (eccentricity of $\sim 0.5$, see Figure~\ref{images}). Unlike most of the other synchrotron-dominated remnants, there is no gamma-ray detection of G330.2+1.0. A H.E.S.S. TeV upper limit is given in \citet{abramowski14}. Radio emission from the remnant was first discovered by \citet{clark75}, with subsequent imaging by \citet{caswell83} and \citet{whiteoak96}. The X-ray spectrum was first described by \citet{torii06}, who used {\it ASCA} data to show that the integrated X-ray spectrum is featureless, and reasonably well-fit with a power-law of spectral index 2.8. Little is known about the age, while the distance is estimated to be 5 kpc based on H I studies \citep{mcclure01}. 

\begin{figure}[htb]
\includegraphics[width=8cm]{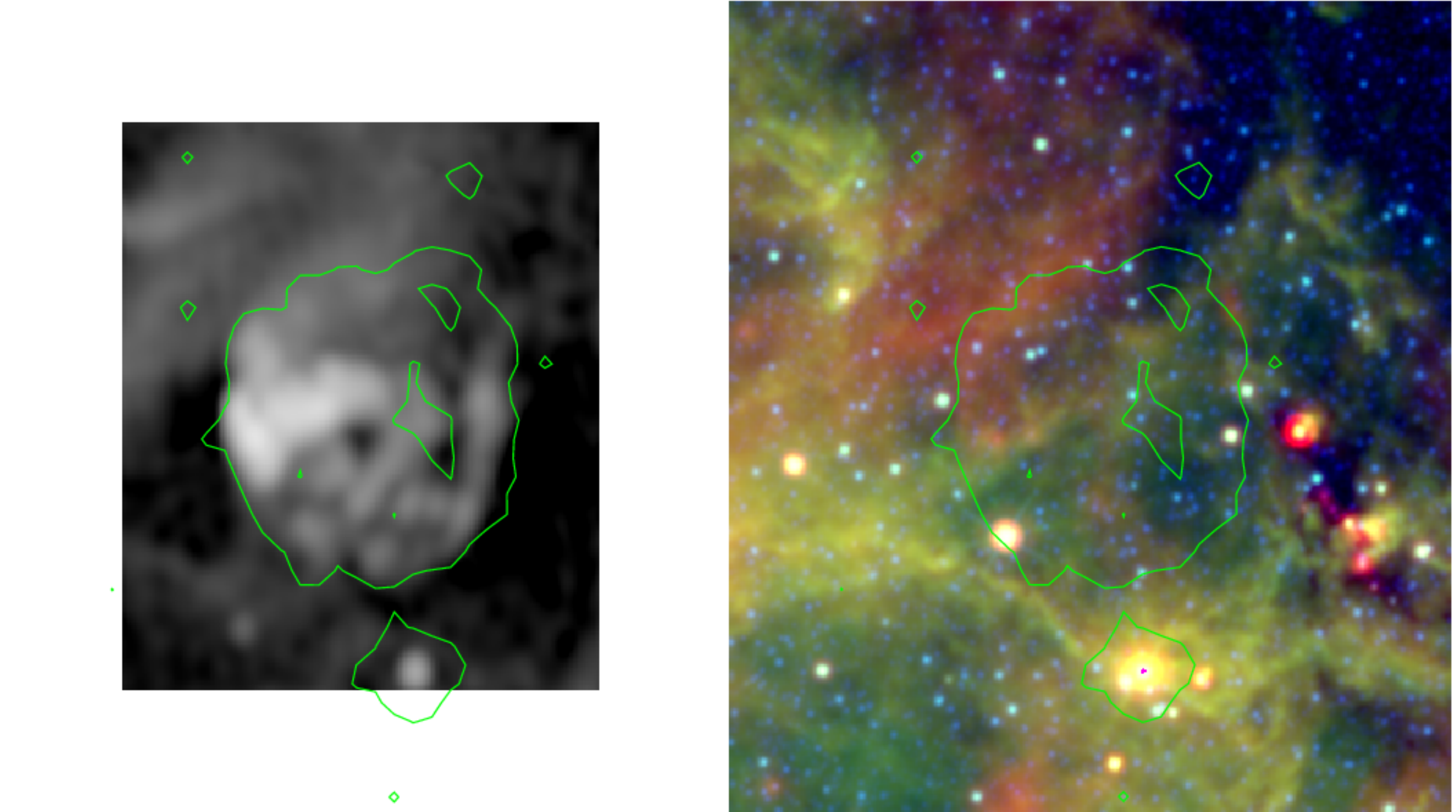}
\caption{{\it Left}: The 843 MHz radio image from \citet{whiteoak96}. {\it Right}: a three-color WISE image, with 22 $\mu$m emission in red, 12 $\mu$m in green, and 4.6 $\mu$m in blue.  The images are on the same scale, a larger scale than Figure~\ref{images}, to show the extended infrared nebulosity in the vicinity of the remnant. A contour of the X-ray emission shown in Figure~\ref{images} is overlaid.
\label{radio_ir}
}
\end{figure}

The 843 MHz radio image of G330.2+1.0 from \citet{whiteoak96} is shown in the left panel of Figure~\ref{radio_ir}. A shell-like structure is visible in most of the remnant, though this feature fades away into the background in the NW. The most striking feature of the radio image is the bright emission in the center of the eastern hemisphere. This emission roughly corresponds to thermal X-ray emission seen in our deep {\it XMM-Newton} images, which we discuss below. A three-color (4.6, 12, and 22 $\mu$m) infrared image from the {\it Wide-Field Infrared Survey Explorer (WISE)} is also shown in Figure~\ref{radio_ir}. SNRs are often seen in the infrared via the emission from warm dust grains in the post-shock environment \citep{williams06}, particularly standing out at 22 $\mu$m. We see no clear emission from G330.2+1.0 in the WISE image. A arc-like feature appears in the vicinity of the NE portion of the X-ray and radio shell, but on taking a wider view, this emission appears to be a part of some unrelated nebulosity extending well away from the remnant. The place where we would most expect to see IR emission, the radio-bright and X-ray thermal region mentioned above, has no emission at all in the WISE image.

The shell is virtually entirely-dominated by pure synchrotron continuum emission in the X-rays, save for a faint thermal region which we will discuss, below. The lack of any thermal emission from the forward shock led P09 to conclude that the preshock density in the ISM is $\sim 0.1$ cm$^{-3}$. The lack of depth of the observations prevented P09 from doing detailed spatially resolved studies of the nonthermal emission, limiting them to only two large regions. These regions were dominated by power-law continua with spectral indices, $\Gamma$, of 2.13 and 2.52, each with substantial uncertainties. They report the detection of faint thermal emission from the eastern lobe, but the low signal-to-noise prevented a detailed characterization of this emission. In this paper, we expand upon this work using deeper observations with an order of magnitude increase in the count rate from the remnant. We do a systematic search for spatial variations in the spectral properties of the nonthermal emission, finding virtually none. We also characterize the thermal emission in the eastern lobe, finding it to be consistent with the forward shock encountering interstellar material. 

\section{Observations}
\label{obs}

G330.2+1.0 was first observed by {\it XMM-Newton} in 2008 for 70 ks. This data was analyzed in P09, where significant background flaring reduced the effective exposure times for the MOS1 and MOS2 detectors to 31 and 33 ks, respectively. The pn observations were taken in small-window mode to search for pulsations in the CCO, and none of the extended emission from the SNR shell falls on the pn detector. Our observations (ObsID 0742050101), taken on 8-9 Mar 2015, totaled 125 ks, with all detectors in full-window mode. After filtering the data for flaring, we were left with approximately 88, 95, and 75 ks for the MOS1, MOS2, and pn detectors. Between the factor of $\sim 3$ greater observing times for MOS1 and MOS2 and the addition of 75 ks of pn data, we have roughly an order of magnitude more counts in any given region than P09 did. Our data were all processed using version 16.0.0 of the {\it XMM-Newton} Science Analysis System (SAS). Spectra were grouped to a minimum of 15 counts per energy bin.

All spectra are background subtracted using an off-source region to the NE of the remnant, in an area that is covered by all three detectors and is free of point sources. We experimented with several different background subtraction regions, and found no statistically significant difference in any of the fits we report below. G330.2+1.0 was also observed with {\it Chandra} in 2006 for 50 ks, but the signal to noise in these observations is comparable to that of the 2008 {\it XMM-Newton} observations (i.e., about an order of magnitude lower than what we have here). The addition of the {\it Chandra} data to our analysis adds virtually nothing to the statistical constraints of the model fits; thus, we do not include the {\it Chandra} data here. For our spectral analysis, we extract the spectra of the MOS1, MOS2, and pn data separately and perform joint fits in XSPEC v12.9.1. 

\section{Spectral Modeling and Results}

\subsection{Nonthermal Emission from the Shell}

\begin{figure}[htb]
\includegraphics[width=8cm]{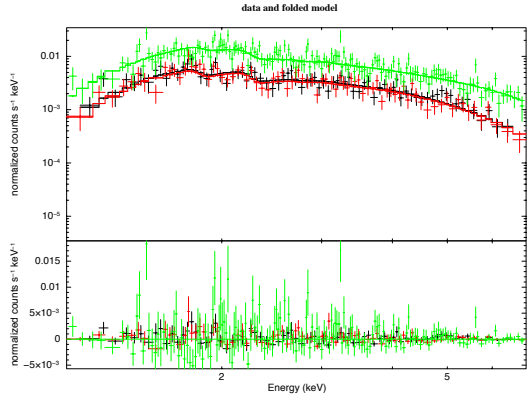}
\caption{The spectrum of a sample region, region 8, with MOS1 in black, MOS2 in red, and pn in green. A model fit of an absorbed power-law is overlaid. The reduced $\chi^{2}$ value of this fit is 0.96.
\label{region8}
}
\end{figure}

The shell of G330.2+1.0 is dominated almost entirely by nonthermal X-rays that exhibit pure continuum emission. P09 only had sufficient photon counts to study two regions, one in the NE and one in the SW. These two regions are best fit by power-laws with indices 2.52 (upper and lower limits of 2.92 and 1.98, respectively) and 2.13 (2.37 and 1.91). With better signal-to-noise, a primary goal of our work here is to investigate whether spatial variations really do exist within the remnant. Such variations could be caused by variations in the shock velocity or the compression of the magnetic field behind the shock \citep{tran15}, or by differences in the angle between the shock front and the Galactic magnetic field \citep{west17}. We discuss this more in Section~\ref{discussion}. To study the spatial variation of the spectral shape, we broke the rim up into fourteen analysis regions distributed azimuthally around the periphery of the shock front, ensuring that each region contained sufficient ($\gtrsim 4000$) counts between the three instruments to fit appropriate spectral models. We also took care to avoid any bright point sources visible in the image. These regions are shown in Figure~\ref{images}. The spectrum of a sample region (region 8) from all three {\it XMM-Newton} detectors is shown in Figure~\ref{region8}.

To fit the spectra, we first took the simplest reasonable approach: the fitting of an absorbed pure power-law. For these initial fits, we fixed the absorbing column density to the value reported in \citep{park06}, 2.5 $\times$ 10$^{22}$ cm$^{-2}$. We jointly fit the spectra from the three instruments with the power-law index, $\Gamma$, as the only free parameter (not including the overall normalization, which was of course allowed to vary from region to region as well). We performed the fits in the energy range of 1$-$7 keV. The values of $\Gamma$ from the spectral fits are reported in Table~\ref{resultsplfix}, along with the goodness of fit, determined by the $\chi^{2}$ value, and the azimuthal angle of the regions, defined as east of north. While not exactly centered in the remnant, we use the compact object as the reference point for the position angle, and measure the angle to the center of the region.

\begin{figure}[htb]
\includegraphics[width=8cm]{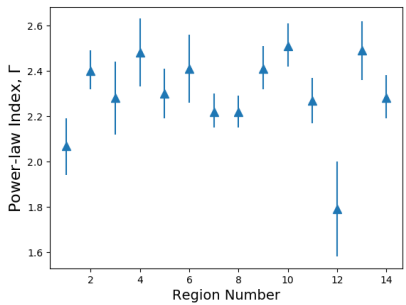}
\caption{Measured values of the power-law index, $\Gamma$, as jointly fit to the MOS1, MOS2, and pn spectra of the 14 regions around the shell of G330.2+1.0. The absorbing column density was fixed at 2.5 $\times 10^{22}$ cm$^{-2}$ for these fits.
\label{plfix}
}
\end{figure}

In Figure~\ref{plfix}, we plot the value of $\Gamma$ as a function of the azimuthal angle, reported by region number. No obvious trends stand out. The average value of 2.29 is nearly exactly in between the values reported for the two regions of P09. The standard deviation of the values is rather small, at 0.18, and if the particularly low value for region 12 ($\Gamma = 1.79$) is ignored, the average and standard deviation for the other 13 regions become 2.33 and 0.12, respectively. We have no {\it a priori} reason to ignore region 12, though, and it is interesting to note that region 12 is right next to one of the highest regions measured, region 13. It is also worth noting that from a purely statistical point of view, the fits to region 12 have the lowest value of $\chi^{2}$ of any region.

\begin{figure}[htb]
\includegraphics[width=8cm]{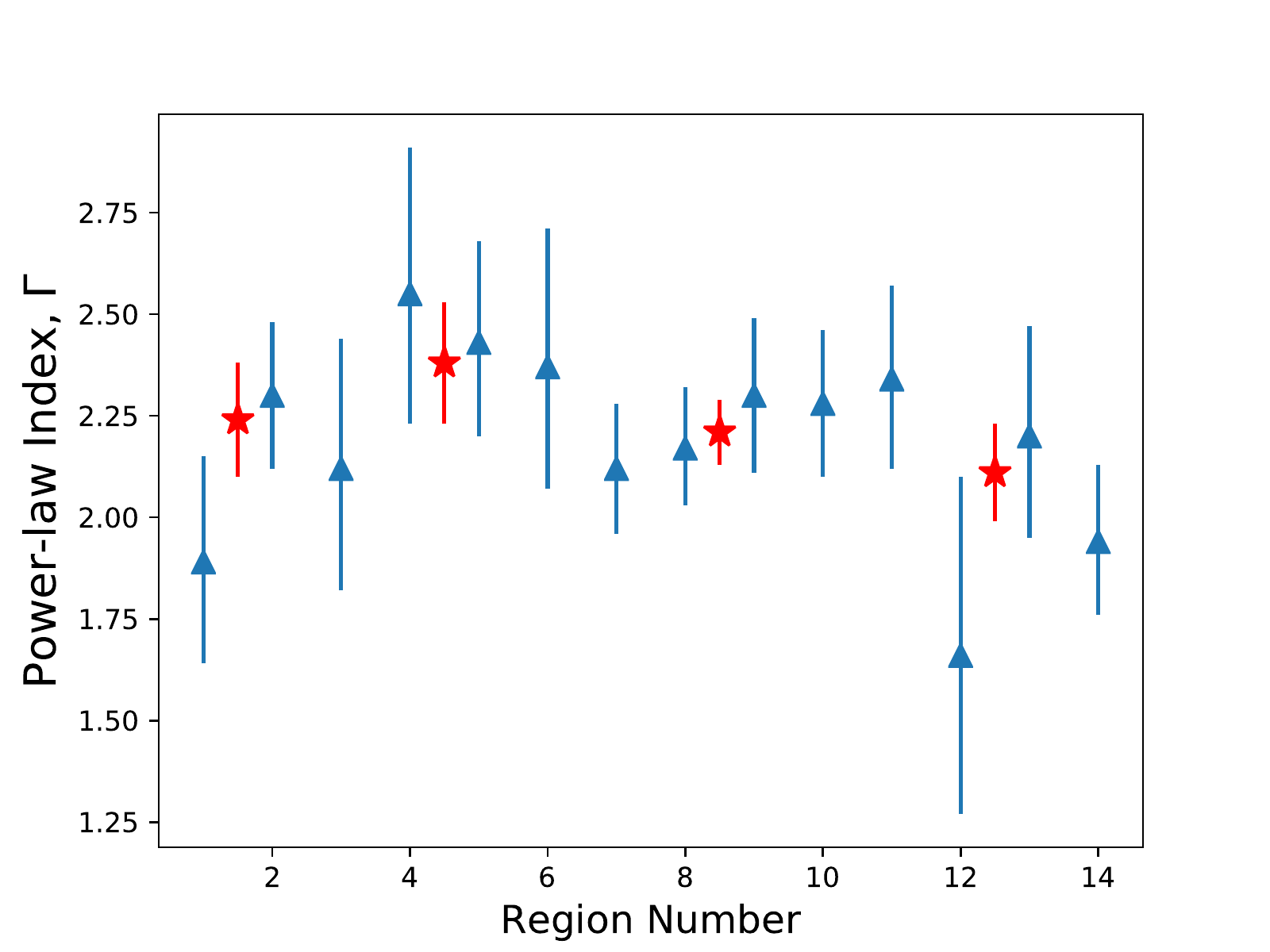}
\caption{Same as Figure~\ref{plfix}, except that the value of the absorbing column density, N$_{H}$, were allowed to float freely. The values for N$_{H}$ are shown in Figure~\ref{nhfree}. Blue triangles mark the fitted values for our 14 regions; red stars mark the values for our four "large" regions, described in the text.
\label{plfree}
}
\end{figure}

\begin{figure}[htb]
\includegraphics[width=8cm]{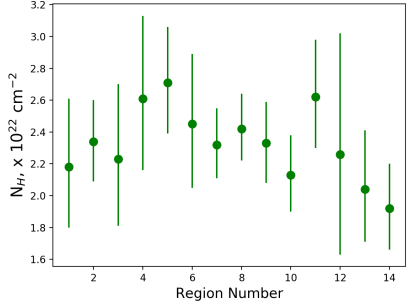}
\caption{The measured values of the absorbing column density for the 14 regions, as jointly fit to the spectra.
\label{nhfree}
}
\end{figure}

As a next step, we investigated the results of the fitting if the absorbing column density for each region was also allowed to float freely. As in the case of fitting the power-law index, the values from the three instruments are tied together and fit jointly. The results are shown in Figure~\ref{plfree} and Table~\ref{resultsplfree}. Again, no obvious trend emerges, as the values are quite similar to the values for the case with the column density fixed. This is not surprising, as the region that we are fitting, 1$-$7 keV, is relatively unaffected by the absorption of low-energy X-rays. The individual values of N$_{H}$ themselves are shown in Figure~\ref{nhfree}. The average value is 2.33 $\times 10^{22}$ cm$^{-2}$, with a standard deviation of 0.22 $\times 10^{22}$ cm$^{-2}$, within errors of the global value of 2.5 $\times 10^{22}$ cm$^{-2}$ from \citet{park06}.

We then repeat the fits with a different model. The {\it SRCUT} model of \citet{reynolds98} describes the synchrotron emission from a power-law distribution of shock-accelerated electrons, with the electron spectrum cutting off as exp($-E/E_{m}$) (and the corresponding photon spectrum cutting off as exp[-($\nu / \nu_{m})^{1/2}$], where $E_{m}$ and $\nu_{m}$ are cutoff frequencies corresponding to the maximum energy to which the shock can accelerate an electron and the frequency of the photon emitted. The parameters for the {\it SRCUT} model are the spectral index of the radio emission, $\alpha$, the break frequency in the photon spectrum, $\nu_{m}$, and the overall normalization. As was done in P09, we fix the radio spectral index, $\alpha$, to -0.3, the global value reported for the remnant in \citet{clark75}. We freeze the value of the absorbing column density to the average value for the remnant determined above, 2.3 $\times 10^{22}$ cm$^{-2}$ (though we note, again, that this has virtually no effect on the fits). 

\begin{figure}[htb]
\includegraphics[width=8cm]{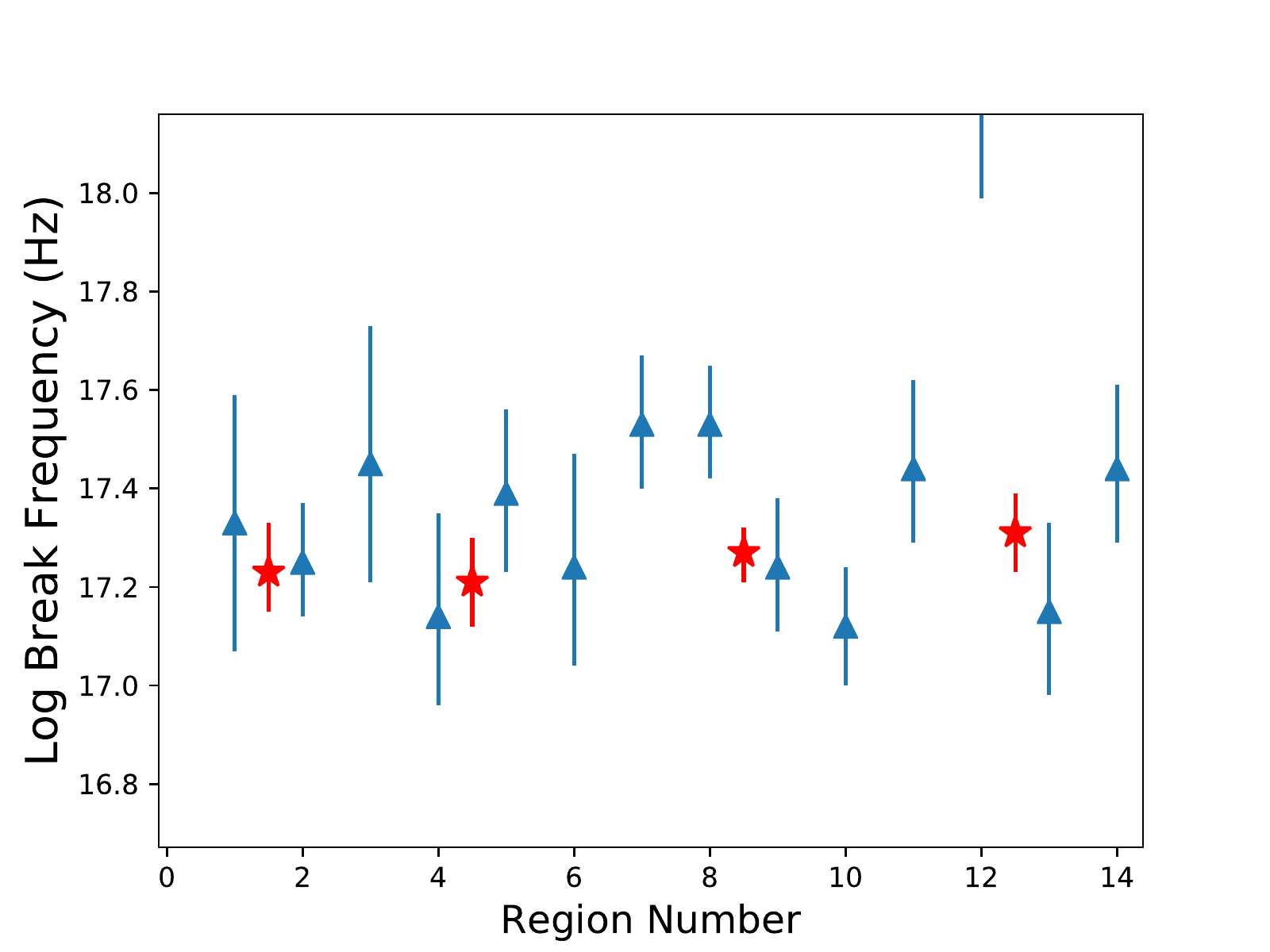}
\caption{The measured values of the spectral break frequency in the {\it SRCUT} model for the 14 regions, as jointly fit to the spectra. The value for region 12 (18.62) is off the top of the chart to allow for easier viewing of the variations of the other 13 regions. Blue triangles mark the fitted values for our 14 regions; red stars mark the values for our four "large" regions, described in the text.
\label{srcutfits}
}
\end{figure}

The results are shown in Figure~\ref{srcutfits} and given in Table~\ref{resultssrcut}. As with the power-law fits above, no systematic trend in the results can be identified. The values for $\nu_{m}$ generally fall between 17.1 and 17.5 Hz in logarithmic space, with only region 12 again as an outlier. The average value, including all regions, is 17.42 Hz with a standard deviation of 0.36. If region 12 is excluded, the average becomes 17.33 Hz, with the standard deviation dropping to 0.14. These results are in stark contrast to what is seen in SN 1006, where \citet{miceli09} show, in Figure 3 of their paper, that the break frequency varies systematically as one moves around the periphery of the shell by more than an order of magnitude.

As a final step, we repeat the fits from the last two models, which we consider to be the most realistic physical scenarios, for four larger regions around the shell. Our "large" regions approximately break the remnant up into quadrants, with regions 1 and 2 forming the NE, regions 3-6 forming the SE, regions 7-10 forming the SW, and regions 11-14 forming the NW. We extract the spectra from these combined regions and fit them as before, to see if any significant variation is present on large scales. We find no such variation is present at a significant level, particularly when compared to remnants like SN 1006. Figure~\ref{plfree} and Figure~\ref{srcutfits} show the fits to these large regions as red stars.

\subsection{Nonthermal Emission from the Interior}

A noticeable feature of G330.2+1.0, as can be seen in Figure~\ref{images}, is the presence of diffuse emission inside the outer rim in the interior of the remnant. This is seen in some SNRs; RCW 86, for instance, contains very similar diffuse nonthermal emission in the interior of the remnant \citep{williams11}. We divided the internal emission up into three distinct regions, fitting each with the same methods as above. Interestingly, the spectra are virtually identical to those from the outer rim. Internal regions 1, 2, and 3 have power-law indices of 2.25, 2.32, and 2.39, respectively. The values of the break frequency in the {\it SRCUT} model are 17.50, 17.39, and 17.15 Hz, respectively. We discuss this further, below.

\subsection{Thermal Emission in the East}

\begin{figure}[htb]
\includegraphics[width=8cm]{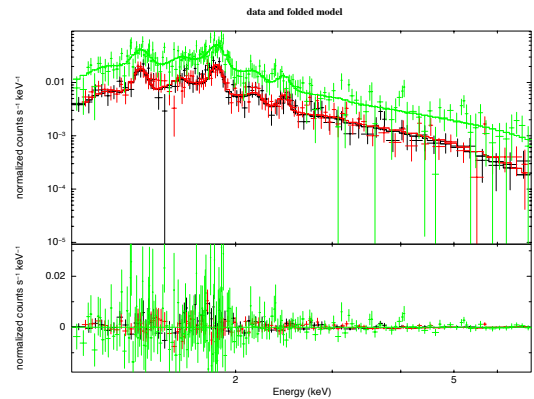}
\caption{The 1-7 keV spectra of the "therm" region in the eastern part of the shell, with MOS1 in black, MOS2 in red, and pn in green. The model described in the text is overlaid. The reduced $\chi^{2}$ value for this fit is 0.97.
\label{thermalspec}
}
\end{figure}

The final feature that stands out the most in Figure~\ref{images} is the green region located along the eastern periphery of the shell. Green in this context is emission from 1.2$-$2.0 keV, in the region of the spectrum where Mg ($\sim 1.35$ keV) and Si ($\sim 1.8$ keV) lines are most prominent. The spectrum from this "therm" region, shown in Figure~\ref{thermalspec} confirms that these lines are indeed present in the spectrum, a clear indication that the emission here is at least partly thermal in nature from a hot, shocked plasma. We obtain quite a good fit (a reduced $\chi^{2}$ value of 0.97 for 452 degrees of freedom) for a simple model consisting of an absorbed power-law plus a non-equilibrium ionization component at normal galactic abundances, with respect to \citet{wilms00} abundances. The underlying power-law parameters were allowed to float freely, settling on a spectral index of 2.15, consistent with the rest of the rim. The ionization timescale in this model was around 3.5 $\times 10^{11}$ cm$^{-3}$ s, while the temperature of the plasma is 0.49 keV. When we allowed the abundance to float freely in {\it XSpec}, the fit selected a best fit value of 0.85, but with large uncertainties (from 0.5 to 4.5). 

To investigate this further, we used an absorbed power-law plus variable abundance plane-shock model ({\it vpshock}) and re-fit the spectrum. Because Mg and Si are the only lines clearly visible in the spectrum, we allowed these to float freely and independently, though their fitted values were virtually identical. The final best-fit spectrum had a virtually identical goodness-of-fit (reduced $\chi^{2}$ of 0.97 for 453 degrees of freedom), but the uncertainties on the fitted parameters were much smaller. We list all values in Table~\ref{thermalfits}. The fits are entirely consistent with a moderate velocity shock encountering ISM material at normal abundances.

\section{Discussion}
\label{discussion}

As stated above, we find no evidence for a systematic trend of spectral changes in the nonthermal spectra of G330.2+1.0 as a function of azimuthal angle. Amongst the synchrotron-dominated remnants listed in the introduction, at least some of them do show such a variation. SN 1006, for instance, shows perhaps the strongest variation of all, with only the limbs on the NE and SW quadrants exhibiting nonthermal emission. Within those limbs, there exists even more variation, as the power-law index (or synchrotron break frequency) rises and falls with azimuthal distance from the peak of the emission \citep{miceli09}. See \citet{rothenflug04} and \citet{miceli13} for more on the spectral variations within SN 1006. G1.9+0.3, the youngest remnant in the Galaxy, shows a similar bipolar structure with spectral variations \citep{reynolds08, reynolds09}. 

\citet{reynolds99} calculate the energy, $E_{max}$, at which the electron energy distribution must steepen in a shock wave. This energy is related to both the magnetic field strength, $B$, and the rolloff energy, $E_{rolloff}=h\nu_{rolloff}$ by the equation

\begin{equation}
E_{max} = 120\ (\frac{h\nu_{rolloff}}{1\ \rm keV})^{1/2}\ (\frac{B}{\mu G})^{-1/2}\ \rm{TeV},
\end{equation}

which comes from \citet{lopez15} after correcting for a small numerical error in \citet{reynolds99}. We do not know the magnetic field strength, but the dependence on this quantity is relatively weak. For a "typical" post-shock value of 10 $\mu$G, the average value of $\nu_{rolloff}$ for our regions around the shell (10$^{17.45}$ Hz) corresponds to an $E_{max}$ of 125 TeV, comparable to other remnants in the synchrotron-dominated class. Note that this implies a fast shock wave. Following the prescription of \citet{castro13}, electrons with an energy $E$ in a magnetic field $B$ will emit synchrotron radiation with a peak energy, $h\nu_{peak}$ of

\begin{equation}
h\nu_{peak} = \frac{B_{100}E^{2}}{520}\ \rm keV,
\end{equation}

where $B_{100} = B/100$. Taking the energy we derive above of 125 TeV, we calculate a peak energy of 3 keV. \citet{vink06} show that this peak energy is related to the shock speed by 

\begin{equation}
V_{s} = (2650\ \rm km\ \rm s^{-1})\ (\frac{h\nu_{peak}}{1\ \rm keV})^{1/2}.
\end{equation}

From this, we derive a shock speed of $\sim 4600$ km s$^{-1}$. At a presumed distance of 5 kpc, this velocity would correspond to a proper motion of $\sim 0.2"$ yr$^{-1}$, detectable by {\it Chandra}, which observed the remnant in 2006 and 2017.

The diffuse nonthermal emission in the interior is reminiscent of that seen in RCW 86 \citep{williams11}. In that paper, we attributed this diffuse synchrotron morphology to "relic" emission from electrons accelerated early on in the lifetime of the remnant, when the shocks were much faster. In that case, these electrons would no longer be confined to the rim, as they would have diffused out. This explanation works well for RCW 86, which is believed to be a cavity explosion into a low-density bubble carved out by a pre-SN wind. It is possible that the same explanation could hold here, and that G330.2+1.0 could also be a cavity explosion. However, if this were the case, one would expect the electrons accelerated at the blast wave to have a different spectrum than those relic electrons accelerated long ago, and we do not see evidence for this in G330.2+1.0. One possibility is that the interior diffuse emission is simply the projection along either the front or back side of the remnant of synchrotron emission at the forward shock. 

It is also possible that G330.2+1.0 is actually quite young, perhaps $< 1000$ yrs. In this case, there would not be time for the spectral shapes to change significantly between the time that the relic electrons were shocked and the current epoch of electrons shock at the periphery of the blast wave, particularly given the uncertainties on the power-law fits. There are no records of a historical SN at this location, but this is not a problem. Assuming a relatively bright (M$_{V}$ = -19) SN, the apparent magnitude at 5 kpc would be m = -6.5, but this does not account for extinction. Using the relationship between column density and extinction from \citet{predehl95} of N$_{H}$/A(V) = 1.79 $\times 10^{21}$ cm$^{-2}$ mag$^{-1}$, a column density of 2.3 $\times 10^{22}$ cm$^{-2}$ leads to an extinction of 13 mag, for an actual apparent magnitude of 6.5, highly unlikely to be detected even by the most astute skywatchers of the time.

In general, the lack of thermal emission at the forward shock implies a low density that the forward shock is encountering. P09 estimate a preshock density of  $n_{0} \sim 0.1$ cm$^{-3}$, comparable to that seen in Tycho \citep{hwang02,williams13}, which also lacks thermal X-rays from the forward shock. However, in the eastern lobe, our spectra show that thermal emission is quite clearly present. This thermal emission corresponds reasonably well morphologically with the brightest radio emission observed. This is not unexpected; as we show below, the thermal fits to the region imply that the shock there is encountering a much higher density than elsewhere in the remnant, and thermal X-ray emission and radio synchrotron emission depend on the density \citep{chevalier82}.

However, it is interesting to note one major difference from Tycho: the lack of thermal emission from the ejecta. In this respect, though, it is more like several of the other synchrotron-dominated remnants, like RX J1713.7$-$3946 and G1.9$+$0.3, which have little or no emission from ejecta in the X-rays, indicating that the reverse shock has not yet formed in these remnants. This indicates a young evolutionary state for G330.2+1.0, and further strengthens our hypothesis that the actual age may also be young.

The temperature we derive for the fit to the thermal region, $\sim 0.5$ keV, is fairly typical of shock waves from SNRs that are a few thousand years old. Assuming standard shock jump conditions, this temperature corresponds to a shock velocity of $\sim 650$ km s$^{-1}$. Yet the spectrum also requires a nonthermal component to be present, with a power-law index comparable to that around the rest of the shell. Also, the section of the blast wave where the thermal emission is present is clearly indented slightly from the rest of the shell, consistent with a shock wave that has experienced significant deceleration relatively recently. This is not unlike the situation in the NW rim of SN 1006, where the densities are found to be of order 1 cm$^{-3}$, more than an order of magnitude higher than the rest of the shell and the only place where thermal emission is seen from the forward shock \citep{winkler13}.

The most straightforward explanation for this is simply that while most of the blast wave is encountering very low density material, possibly carved out by a pre-SN wind, a small section is beginning to encounter a denser region of either the ISM or a clump of circumstellar material (CSM). The abundances we find in this region are entirely consistent with ISM densities. In addition, we can derive a rough estimate of the density from the emission measure of the shocked gas as fit by our {\it vpshock} model. Our best fit emission measure is 4.9 $\times 10^{-3}$ cm$^{-5}$. The area of the "therm" region is approximately 200" $\times$ 120". If we make the ballpark assumption that the region is as deep as the average of these two values (or 160"), we arrive at a volume 1.6 $\times 10^{57}$ cm$^{-3}$. If we assume a distance of 5 kpc, we derive a density of the emitting material of $\sim 1$ cm$^{-3}$, the approximate density of the galactic ISM. There is clearly some uncertainty in this number: in addition to the assumptions about the depth of the emitting region and the distance to the remnant, we also assume a filling fraction of unity, so this derived density is likely only accurate to a factor of a few. Still, that it falls so close to expected ISM densities reinforces our hypothesis.

One might expect to see some thermal IR emission from warm dust grains in the post-shock plasma, but as mentioned above, the WISE All-Sky Survey images show no obvious emission. This is likely due to a confluence of factors. First, the IR emission in that part of the sky (G330.2+1.0 is only one degree out of the Galactic plane) is dominated by diffuse galactic emission, as Figure~\ref{radio_ir} shows. Secondly, WISE is not particularly sensitive to SNRs, as it only goes up to 22 $\mu$m. With a post-shock density of around 1 cm$^{-3}$, any dust here would not be very hot; as compared with, e.g., Kepler's SNR \citep{williams12}. The dust temperatures produced by such densities would cause grains to emit more at longer wavelengths, more suitable for the more sensitive {\it Spitzer} 24 and 70 $\mu$m bands. Even then, the emission would be difficult to disentangle from the high background levels. Unfortunately, this remnant was never observed with the cryogenic instruments on {\it Spitzer}.

Although virtually all of our regions are statistically consistent with a pure power-law (or {\it srcut}) model fit, we re-fit the data for all 14 regions around the periphery of the rim using a power-law plus a thermal component. We did this for two reasons: to see whether any faint thermal emission was present anywhere else in the remnant, and to put upper limits on the density of the material the forward shock is encountering. In none of the 14 regions did the addition of a thermal component make any statistical difference to the fits. The upper limits to the density that we derived were generally very low. As an example, we consider region 8 (shown in Figure~\ref{region8}). We added a thermal model to the power-law fit, dialing up the emission measure of the thermal component until it began to affect the statistical goodness of the fit. We fit a value of the emission measure (in {\it Xspec} units of "norm") of 8 $\times 10^{-6}$ cm$^{-5}$ for this region. Using the same technique as above of translating this to a density, we derive an upper limit to the density of 0.11 cm$^{-3}$, consistent with the value from P09. 

As a caveat to this, we point out that in doing these "upper limit" fits, we froze the value of the temperature and ionization timescale of the plasma to that fit for the thermal region reported in Table~\ref{thermalfits}, and the abundances to their cosmic values. Failure to do so would result in unconstrained fits (i.e., one could theoretically fit virtually any value for the emission measure, and thus any value of the density, if, say, the abundances or the ionization timescale were particularly low). The upper limits to the density that we derived from this method were quite consistent throughout the remnant, varying by less than a factor of $\sim 50\%$ from the 0.11 cm$^{-3}$ calculated above. Even in region 13, which had the worst statistical fit with the nonthermal emission models only ($\chi^{2} \sim 1.25$), the addition of a thermal component did not improve the statistical quality of the fit, and the upper limit derived using this method resulted in a density value of 0.08 cm$^{-3}$.

\section{Conclusions}

We have presented the deepest X-ray view yet of SNR G330.2+1.0, a remnant dominated by nonthermal synchrotron emission. Our 125 ks observations collected an order of magnitude more photon counts than previous {\it XMM-Newton} observations, allowing us to probe the nature of the emission on much smaller scales. We examined 18 total regions: 14 from around the periphery of the forward shock, three in the interior, and one from a region that also showed thermal emission. Our conclusions are as follows:

- Variations do exist in the spectral properties (power-law index, roll-off frequency, or absorbing column density) for the various regions. However, the error bars generally overlap with each other, and we find no systematic trend of variation with the location within the remnant. Such trends are sometimes seen in synchrotron-dominated remnants, with SN 1006 being a prime example.

- The fits to the spectra show rather high values for the roll-off frequency of the synchrotron emission (an average value of 10$^{17.45}$ Hz), implying high values of $E_{max}$ of $\sim 125$ keV. This also implies fast shocks of 4600 km s$^{-1}$ on average. At a distance of 5 kpc, this would lead to a proper motion of $\sim 0.2"$ yr$^{-1}$. 

- The nonthermal emission from the interior has virtually identical spectral properties to that at the forward shock. This implies that either this interior emission is simply emission from the front or back side of the remnant, seen in projection (and thus, still coincident with the forward shock), or that the remnant is quite young, perhaps $<$ 1000 yrs.

- The thermal emission in the east is well-fit by a shock model encountering a region of ISM/CSM at normal cosmic abundances. Fits yield a temperature of about 0.5 keV and an ionization timescale around 5 $\times 10^{11}$ cm$^{-3}$ s. Reasonable estimates for the volume of the region imply a density of 1 cm$^{-3}$.

- Elsewhere in the remnant, there is no hint of thermal emission from the shock front. We place upper limits on the density of the ISM/CSM for most of the remnant at $< 0.1$ cm$^{-3}$, consistent with the fast shock waves derived from the synchrotron spectral fits.

\begin{deluxetable}{lccccccc}
\tablecolumns{7} 
\tablewidth{0pc} 
\tabletypesize{\footnotesize}
\tablecaption{Power-law Fits, N$_{H}$ = 2.5 $\times 10^{22}$ cm$^{-2}$} \tablehead{ \colhead{Region} & Deg & $\Gamma$ & Low & High & $\chi^{2}$ & d.o.f. & Reduced $\chi^{2}$}

\startdata

1 & 35.7 & 2.07 & 1.94 & 2.19 & 151.5 & 163 & 0.95\\
2 & 58.4 & 2.40 & 2.32 & 2.49 & 291.1 & 254 & 1.16\\	
3 & 139.1 & 2.28 & 2.12 & 2.44 & 226.3 & 188 & 1.20\\
4 & 165.6 & 2.48 & 2.33 & 2.63 & 135.4 & 128 & 1.09\\
5 & 183.8 & 2.30 & 2.19 & 2.41 & 194.9 & 198 & 0.98\\
6 & 200.8 & 2.41 & 2.26 & 2.56 & 105.9 & 109 & 0.97\\
7 & 219.1 & 2.22 & 2.15 & 2.30 & 275.0 & 273 & 1.01\\
8 & 236.5 & 2.22 & 2.15 & 2.29 & 305.5 & 318 & 0.96\\
9 & 252.2 & 2.41 & 2.32 & 2.51 & 256.7 & 265 & 0.97\\
10 & 271.8 & 2.51 & 2.42 & 2.61 & 300.9 & 256	 & 1.18\\
11 & 292.1& 2.27 & 2.17 & 2.37	 & 285.3 & 262 & 1.09\\
12 & 302.6 & 1.79 & 1.58 & 2.00 & 87.2 & 99 & 0.88\\
13 & 327.4 & 2.49 & 2.36 & 2.62 & 303.7 & 238	 & 1.28\\
14 & 344.0 & 2.28 & 2.19 & 2.38 & 267.0 & 275	 & 0.97\\

\enddata

\tablecomments{Deg. = Position Angle, measured E of N. $\Gamma$ = power-law index. Low, High = lower and upper limits, respectively. d.o.f. = degrees of freedom in spectral fit}
\label{resultsplfix}
\end{deluxetable}

\newpage
\clearpage

\begin{deluxetable}{lcccccccccc}
\tablecolumns{10} 
\tablewidth{0pc} 
\tabletypesize{\footnotesize}
\tablecaption{Power-law Fits, N$_{H}$ free} \tablehead{ \colhead{Region} & Deg & $\Gamma$ & Low & High  & N$_{H}$ & Low & High & $\chi^{2}$ & d.o.f. & Reduced $\chi^{2}$}

\startdata

1 & 35.7 & 1.89 & 1.64 & 2.15 & 2.18 & 1.80 & 2.61 & 149.9 & 158 & 0.95\\
2 & 58.4 & 2.30 & 2.12 & 2.48 & 2.34 & 2.09 & 2.60 & 290.1 & 249 & 1.16\\
3 & 139.1 & 2.12 & 1.82 & 2.44 & 2.23 & 1.81 & 2.70 & 225.4 & 187 & 1.21\\
4 & 165.6 & 2.55 & 2.23 & 2.91 & 2.61 & 2.16 & 3.13 & 135.2 & 123 & 1.10\\
5 & 183.8 & 2.43 & 2.20 & 2.68 & 2.71 & 2.39 & 3.06 & 193.9 & 197 & 0.98\\
6 & 200.8 & 2.37 & 2.07 & 2.71 & 2.45 & 2.05 & 2.89 & 105.8 & 108 & 0.98\\
7 & 219.1 & 2.12 & 1.96 & 2.28 & 2.32 & 2.11 & 2.55 & 273.3 & 272 & 1.00\\
8 & 236.5 & 2.17 & 2.03 & 2.32 & 2.42 & 2.22 & 2.64 & 305.2 & 317 & 0.96\\
9 & 252.2 & 2.30 & 2.11 & 2.49 & 2.33 & 2.08 & 2.59 & 255.5 & 264 & 0.97\\
10 & 271.8 & 2.28 & 2.10 & 2.46 & 2.13 & 1.90 & 2.38 & 295.2 & 255 & 1.16\\
11 & 292.1 & 2.34 & 2.12 & 2.57 & 2.62 & 2.30 & 2.98 & 284.9 & 261 & 1.09\\
12 & 302.6 & 1.66 & 1.27 & 2.10 & 2.26 & 1.63 & 3.02 & 86.9 & 98 & 0.89\\
13 & 327.4 & 2.20 & 1.95 & 2.47 & 2.04 & 1.71 & 2.41 & 299.6 & 237 & 1.26\\
14 & 344.0 & 1.94 & 1.76 & 2.13 & 1.92 & 1.66 & 2.20 & 256.6 & 274 & 0.94\\

\enddata

\tablecomments{Same as Table~\ref{resultsplfix}, with additional columns for the fitted column density, N$_{H}$ and the lower and upper limits, Low and High.}
\label{resultsplfree}
\end{deluxetable}

\newpage
\clearpage

\begin{deluxetable}{lccccccc}
\tablecolumns{7} 
\tablewidth{0pc} 
\tabletypesize{\footnotesize}
\tablecaption{{\it SRCUT} fits, N$_{H}$ = 2.3 $\times 10^{22}$ cm$^{-2}$} \tablehead{ \colhead{Region} & Deg & log$_{10}$ ($\nu_{rolloff}$) & Low & High & $\chi^{2}$ & d.o.f. & Reduced $\chi^{2}$}

\startdata

1 & 35.7 & 17.33 & 17.07 & 17.59 & 153.1 & 159 & 0.96\\
2 & 58.4 & 17.25 & 17.14 & 17.37 & 295.8 & 250 & 1.18\\
3 & 139.1 & 17.45 & 17.21 & 17.73 & 228.4 & 188 & 1.21\\
4 & 165.6 & 17.14 & 16.96 & 17.35 & 138.9 & 125 & 1.11\\
5 & 183.8 & 17.39 & 17.23 & 17.56 & 196.6 & 198 & 0.99\\
6 & 200.8 & 17.24 & 17.04 & 17.47 & 108.9 & 109 & 1.00\\
7 & 219.1 & 17.53 & 17.40 & 17.67 & 280.6 & 273 & 1.03\\
8 & 236.5 & 17.53 & 17.42 & 17.65 & 308.2 & 318 & 0.97\\
9 & 252.2 & 17.24 & 17.11 & 17.38 & 263.8 & 265 & 1.00\\
10 & 271.8 & 17.12 & 17.00 & 17.24 & 307.7 & 256 & 1.20\\
11 & 292.1 & 17.44 & 17.29 & 17.62 & 285.6 & 262 & 1.09\\
12 & 302.6 & 18.62 & 17.99 & 20.04 & 87.55 & 99 & 0.88\\
13 & 327.4 & 17.15 & 16.98 & 17.33 & 308.7 & 238 & 1.30\\
14 & 344.0 & 17.44 & 17.29 & 17.61 & 273.3 & 275 & 0.99\\

\enddata

\tablecomments{Same as Table~\ref{resultsplfix}, but with the log of the break frequency, $\nu_{rolloff}$, as fit from the {\it SRCUT} model. For the fits, $\alpha$, the radio spectral index, was fixed to 0.3.}
\label{resultssrcut}
\end{deluxetable}

\newpage
\clearpage

\begin{deluxetable}{lccc}
\tablecolumns{3} 
\tablewidth{0pc} 
\tabletypesize{\footnotesize}
\tablecaption{Parameters of {\it vpshock} fit to "therm" region} \tablehead{ \colhead{Parameter} & Value & Low & High}

\startdata

N$_{H}$ ($\times 10^{22}$ cm$^{-2}$)  & 2.63 & 2.37 & 2.99\\
$\Gamma$ & 2.06 & 1.59 & 2.48\\
kT (keV) & 0.46 & 0.40 & 0.58\\
Mg & 0.77 & 0.65 & 1.03\\
Si & 0.76 & 0.58 & 1.07\\
$\tau_{i}$ ($\times 10^{11}$ cm$^{-3}$ s) & 4.76 & 2.10 & 53.7\\
norm ($\times 10^{-3}$ cm$^{-5}$) & 4.93 & 2.37 & 9.10\\

\enddata

\tablecomments{Abundances for all elements other than Mg and Si are fixed to 1.}
\label{thermalfits}
\end{deluxetable}

\end{document}